\pdfoutput=1
\documentclass{article}
\usepackage{stackengine}
\usepackage{caption}
\newcommand{\etal}{\textit{et al}.}

\usepackage{nips}

\usepackage{graphicx}
\usepackage[utf8]{inputenc} % allow utf-8 input
\usepackage[T1]{fontenc}    % use 8-bit T1 fonts
\usepackage{hyperref}       % hyperlinks
\usepackage{url}            % simple URL typesetting
\usepackage{booktabs}       % professional-quality tables
\usepackage{amsfonts}       % blackboard math symbols
\usepackage{nicefrac}       % compact symbols for 1/2, etc.
\usepackage{microtype}      % microtypography
\usepackage{mathtools} %loads amsmath as well
\usepackage{subfig}

\title{Detecting GAN generated errors}
\author{%
  Xiru Zhu*\\
  Department of Computer Science\\
  McGill University\\
  \texttt{xiru.zhu@mail.mcgill.ca} \\
  \And
  Fengdi Che*\\
  Department of Computer Science\\
  McGill University\\
  \texttt{fengdi.che@mail.mcgill.ca} \\
  \And
  Tianzi Yang \\
   Department of Computer Science\\
  McGill University\\
  \texttt{tianzi.yang@mail.mcgill.ca} \\
    \And
  Tzu-Yang Yu \\
  Department of Computer Science\\
  McGill University\\
  \texttt{tzu-yang.yu@mail.mcgill.ca} \\
  \And
  David Meger \\
  Department of Computer Science\\
  McGill University\\
  \texttt{david.meger@mcgill.ca}\\
  \And
  Gregory Dudek \\
  Department of Computer Science\\
  McGill University\\
  \texttt{gregory.dudek@mcgill.ca}\\
}

\begin{document}
\thanks{First two authors have equal contribution}
\maketitle

\begin{abstract}
Despite an impressive performance from the latest GAN for generating hyper-realistic images, GAN discriminators have difficulty evaluating the quality of an individual generated sample. This is because the task of evaluating the quality of a generated image differs from deciding if an image is real or fake. A generated image could be perfect except in a single area but still be detected as fake. Instead, we propose a novel approach for detecting where errors occur within a generated image. By
collaging real images with generated images, we compute for each pixel, whether it belongs to the real distribution or generated distribution. Furthermore, we leverage attention to model long-range dependency; this allows detection of errors which are reasonable locally but not holistically. For evaluation, we show that our error detection can act as a quality metric for an individual image, unlike FID and IS. We leverage Improved Wasserstein, BigGAN, and StyleGAN to show a ranking based
on our metric correlates impressively with FID scores. Our work opens the door for better understanding of GAN and the ability to select the best samples from a GAN model.             
\end{abstract}

% \begin{figure}
%   \centering
%   \includegraphics[width=14cm]{2-3-03.png}
%   \\ Images sampled from the worst images compared to the best images ranked by PD. Our model detects both quality issues and mode collapse. 
%   \label{Fig1}
% \end{figure}

\section{Introduction}
Recent developments in Generative Adversarial Networks (GAN) have showed rapid improvements in generated image quality and variety. GAN can now generate both ImageNet at reasonable quality and human faces at resolution of 1024 by 1024 \cite{brock2018large, karras2018style}. Beyond generating new data, GAN's widespread usage ranges from inpainting, image super-resolution, image to image translation and learning policy for imitation learning \cite{isola2017image, zhu2017unpaired, yeh2017semantic, yang2017high, Liu_2018_ECCV, ledig2017photo, sonderby2016amortised, ho2016generative}. Despite impressive results, GAN generated images are difficult to analyze. Current benchmarks for GAN performance consist of Inception Score (IS) and the Fretchet Inception Distance (FID). FID and IS are adequate for evaluating overall GAN performance for a large dataset. However, they cannot evaluate the quality of a single image. Furthermore, FID and IS are difficult to interpret; we do not know where unrealistic elements occur.
\begin{figure}
  \centering
  \includegraphics[width=14cm]{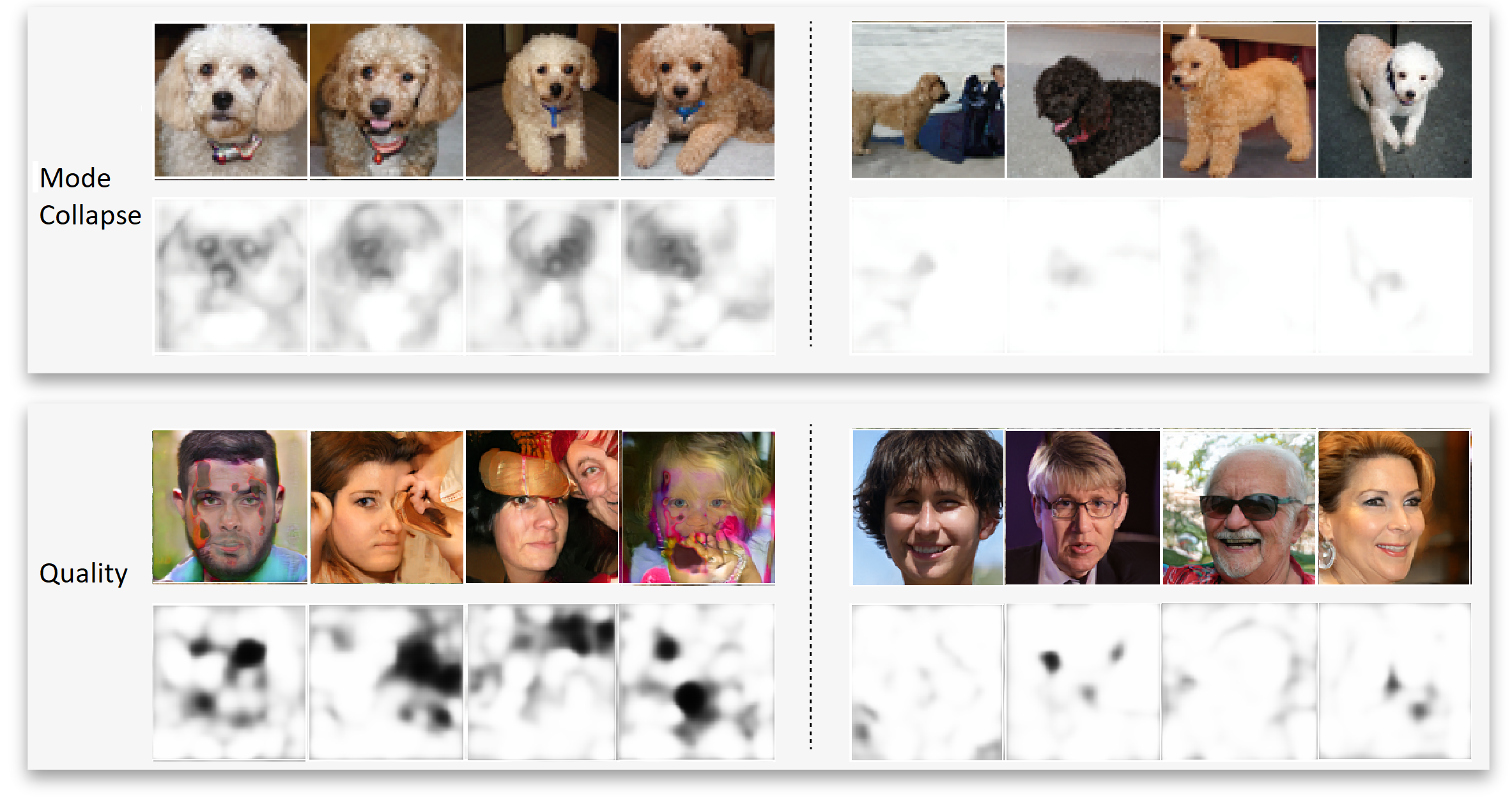}
  \label{fig1}
  \caption{Our approach can detect both mode collapse and image quality problems at the pixel level. The generated dogs from BigGAN are of high quality but suffer from mode collapse around the face. The less mode collapse samples are more representative of the real data distribution is much more diverse. In contrast for human faces generated from StyleGAN, we can see serious artifacts from the worst samples. Our model shows the location of artifacts accurately. 
  }
\end{figure}
When staring at a GAN generated image, one cannot help but to try and find its flaws. Despite researchers' best attempts, state of the art models still generate strange background, weird hats, and other artifacts.
%Bau et al. has shown that a few neurons are responsible for many artifacts in generated images \cite{bau2018gan}. Thus, turning off some of the neurons can improve image quality if we can locate the errors. 
Given the shortcoming of FID and IS, we propose a novel supervised approach for detecting errors and ranking GAN generated images. With our approach, we can detect quality and mode collapse errors at the pixel level as shown in Figure 1. To train without manually labeled data, we collage generated and real images together with some artifacts to mimic errors in the generated distribution. This defines a third image distribution $q$. Our method detects if a given pixel is drawn from the real pixel distribution or the generated pixel distribution $g$. Doing so allows us to visualize GAN generated errors. To evaluate our approach, we propose a performance metric for quality and mode collapse using detected errors. We show our metric correlates well with FID and with qualitative tests.

%To label errors, we collage generated and real images together. Unlike a GAN discriminator which is trained on whether an image is real or generated, our approach looks at the pixel level instead. Our approach is unlike a GAN loss of real or generated which does not generalize well to image quality. A GAN discriminator only seeks to detect if an image is real or not. As such, a generated image could score well for FID but can still be detected as fake. In contrast, our model when applied to a generated image can detect which pixels does not fit the real image distribution. Furthermore, our approach can help visualize a generated image's errors unlike a simple score. To evaluate our approach, we use the computed error for each image and create a quality and mode collapse performance metric for GAN. This can automate the tedious task of sorting image by quality and provide detailed information about areas of the image. 

The major contributions of this paper can be summarized as follows:

\begin{enumerate}
\item \textbf{}We explore a novel approach of detecting quality and mode collapse errors from GAN generated images without manually labeled data. Our error detection can provide information for individual images at pixel level. Furthermore, our detected error can be visualized. 

\item \textbf{}We propose a new performance metric for GAN. Unlike FID and IS, our metric can be used on an individual image and is useful for evaluating a small class of images. 
% We propose a second performance metric for detecting mode collapse only. 

\item \textbf{}We show our model's performance samples from the state of the art approach, BigGAN and StyleGAN for ImageNet and Flickr respectively. Furthermore, we test our approach on badly generated samples from an Improved Wasserstein trained on dogs and cats. 

\item \textbf{}We provide detailed analysis on the performance of BigGAN with our proposed performance metric. 

\end{enumerate}
% The remainder of the paper is organized as follows. Section II gives a brief review on the related work. Section III describes in detail our proposed method. Section IV evaluates the performance of the proposed algorithm against state of the art algorithms using FID.
% Section V concludes this paper with future work and other thoughts. 

\section{Related Works}
\subsection{Generative Adversarial Network}A generative adversarial network (GAN) consists of a discriminator and a generator. The goal of the discriminator is to differentiate between real and generated images. In contrast, the generator's goal is to create realistic images which can fool the discriminator. GAN loss proposed by Goodfellow \etal, minimizing the Shannon-Jensen divergence is as follows \cite{goodfellow2014generative}:
\begin{equation}
    \stackunder{min}{G} \; \stackunder{max}{D} 
    \; = E_{x \sim p_{data}(x)}[log(D(x)))] + E_{z \sim p_{z}(z)}[log(1 - D(G(z))]
\end{equation}
 As the generator is trained with guidance from the discriminator, GAN research has focused on improving the discriminator. Early GAN were unstable during training  \cite{goodfellow2014generative,radford2015unsupervised, mao2017least}. As an improvement, Arjovsky \etal $\;$ proposed using earth mover distance instead of Shannon Jensen Divergence  \cite{arjovsky2017Wasserstein}. This led to significant improvements for training stability and greater sample variety. The Improved Wasserstein proposed a soft clipping which improved training speeds \cite{gulrajani2017improved}. Despite stable training, Improved Wasserstein GAN has trouble generating realistic samples for datasets such as ImageNet. The next milestone for GAN is SAGAN which can generate ImageNet at high quality \cite{zhang2018self}. SAGAN leveraged self attention layer which allowed connections between distant pixels and it improved global image quality. Brock \etal $\;$ took SAGAN to its limits by greatly increasing training layers and training data size \cite{brock2018large}. The resulting work, aptly named BigGAN, showed large performance improvements over SAGAN. 
 
 Besides generating ImageNet, another challenge for GAN is generating high resolution images \cite{zhang2017stackgan}. Progressive GAN lead to a major breakthrough, generating images of size 1024 by 1024. As the name implies, progressive GAN trains low resolution images to high resolution images akin to curriculum learning \cite{karras2017progressive}. Its successor, StyleGAN increases variety by leveraging style transfer and adding noise between layers \cite{karras2018style}.

\subsection{GAN Evaluation Metrics}
Given GAN synthesizes novel data, it remains difficult to measure its performance besides using human judgment. To overcome this problem, Salimans \etal $\;$ proposed the inception score (IS) as a performance metric \cite{salimans2016improved}. As a metric, IS rates image quality and diversity in the classes generated for a large dataset. To do so, IS leverages computed class probability from an inception model and calculates the difference between label probability and the marginal probability.  However, IS has weaknesses; it does not perform well for datasets which are not ImageNet \cite{barratt2018note}. In contrast, the Frechet Inception Distance (FID) measures the distance between the real distribution and fake distribution \cite{heusel2017gans}. As with IS, FID also measures quality and sample diversity. FID computes the features from a pretrained inception model and aggregates statistics for comparison. FID removes the weakness of IS when used for datasets beyond ImageNet. However, the weakness of both approach is that they cannot evaluate the performance of an individual sample. Furthermore, neither FID nor IS provides information on why a model did well or poorly. Finally, as a metric for GAN sample diversity, Arora and Zhang proposed the Birthday paradox test. This approach consists of analyzing generated images and finding the size of a dataset needed before duplicate samples appear. However, this approach is qualitative and requires human judgment \cite{arora2017gans}.

\section{Error Detection for GAN}
To discover errors generated by GAN, we must learn the difference between real distribution $r$ and generated distribution $g$ akin to a GAN discriminator. In contrast to GAN discriminators which are trained to classify images as real or fake, our model is trained to detect salient features of generated distribution at pixel level which we refer to as errors. The advantage of our approach is that it provides richer information for evaluating and understanding GAN. Furthermore, a pixel level loss can guide the model to learn finer details. Let $x$ be a sampled image. Let $E$ equals to 0 when no error exists and let E equals to 1 when there exists errors. Furthermore, let $P(E = e|x, i, j)$ be the probability of error conditioned on image $x$ and its pixel location $i, j$.
Thus, an idealized loss would be
\begin{equation}
    Loss \; = \sum_{i,j}^{N, M} E_{x_r \sim r}[P(E = 1|x_r,i,j)] +  E_{x_g \sim g}[P(E = 0|x_g,i,j)]
\end{equation}
Where $N$ and $M$ are the dimensions of the $x$. However, we do not know where errors are located so we cannot train directly using $r$ or $g$. Instead, we create a distribution $q$ which consist of collaged images drawn from both $r$ and $g$. $q$ is meant to mimic $g$ and $r$ but with knowledge of where errors occurs. Here, errors would refer to pixels in the collaged image from $g$.  
As such, we use the following approximation:
\[ T(x_{q, i, j}) =
  \begin{cases*}
    1 \quad& if $x_{q, i, j}$ is from real \\
    0 & if $x_{q, i, j}$ is from generated \\
  \end{cases*}\]
\begin{equation}
    Loss = E_{x_q \sim q}[ \sum_{i,j}^{N, M} 
    \lambda * T(x_{q, i, j}) * (P(E = 1| x_q, i, j)) + \gamma * (1 - T(x_{q, i, j})) * (P(E = 0| x_q, i, j))) ]
\end{equation}
The two terms in the equation can be explained as predicting errors when there are none, weighted by $\lambda4$ and predicting no errors when there are weighted by $\gamma$. 
We split our equation into two terms so that we can weight the loss for predicting a real pixel to be an error. In a realistic generated sample, predicting it as real is acceptable. However, predicting a real sample as generated should not be. Furthermore, we add $L2$ regularization to help the model generalize results. With $P(E=1|x_g, i, j)$ as the computed results for a generated image,  we can define a new performance metric for GAN, averaged Pixel Distance(PD). 
\begin{equation}
    PD_{x_g} = 1/(N * M) * \sum_{i,j}^{N, M} P(E=1|x_g, i, j)
\end{equation}
Averaged Pixel Distance is similar to FID in that it computes both the quality and mode collapse of GAN generated samples. The difference is that it can compute a value for an individual image rather than for a dataset. Furthermore as we have a score for each pixel, we can even compute PD for parts of an image. 
% From PD, we define a second metric for evaluating mode collapse. Given real and generated data, we can compute the variance of the PD of real and generated samples. Without mode collapse, GAN should generate a variety of image, leading to PD with high variance.  With mode collapse, the variance of PD is reduced as it generates similar images. We define Pixel Variance Distance(PVD) for a generated dataset $X_{fake}$ and real dataset $X_{real}$ as follow:
% \begin{equation}
%     PVD = Var(PD(X_{fake})) - Var(PD(X_{real})) 
% \end{equation}
% Where we offset the computed variance with the real data's PD variance to enable comparison between different classes. Otherwise, certain classes are more diverse to begin with and would always have higher variance. 

\subsection{Training Data and Error Labels}
\begin{figure*}[t] 
    \centering
  \subfloat[]{%
      \includegraphics[width=3.5cm]{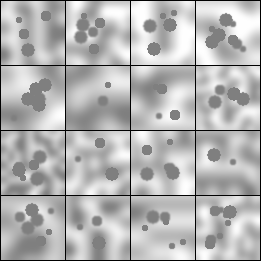}}
  \subfloat[]{%
      \includegraphics[width=3.5cm]{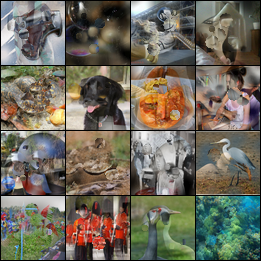}}
  \label{fig2} 
  \caption{
  (a) Perlin Noise with Artifacts.
  (b) Training data from combining generated and real data.}
\end{figure*}

Given we have no knowledge of where error occurs in GAN, we approximate GAN generated image distribution by combining generated images and real images. Our method is different from denoising or image augmentation approaches from past literature  \cite{jain2009natural, xie2012image, warde2016improving}. 
Our goal is not adding noise to improve model robustness but to learn errors. Instead, we collage patches of a generated image to a real image. To randomly combine images, one approach from Liu \etal $\;$ is to create irregular masks \cite{Liu_2018_ECCV}. However, their approach is binary where a pixel is masked or not. Instead, we leverage Perlin Noise from computer graphics\cite{perlin1985image}. Perlin noise is a gradient noise, and is used to generate textures and cloud like patterns. As seen in Figure 2(a), this method can create continuous random noises. Continuity is beneficial as it improves the model's ability to learn a difficult problem. 

Besides merging real and generated images, we further add artifacts using real pixels. To do so, we copy a circular area of pixels from a sampled area on real image. We proceed to replicate these pixels with rotations to a random uniformly sampled area on the real image. The purpose is to create artifacts such as multiple heads or eyes. Such artifacts can help the model detect realistic features at abnormal locations. The circular like artifact can be seen in Figure 2(b). 

\subsection{Network Design}
Our network takes an image of shape $[N,M, 3]$ as input and generates a $PD_{gen_{i,j}}$ value for each pixel to an output shape $[N, M, 1]$. As the output dimension is the similar to the input, we choose use an auto encoder architecture. Our architecture, is based on the U-Net which has skip links connecting the encoder and decoder layer \cite{ronneberger2015u}. Our deep auto-encoder architecture consists of 5 layers of convolution per pooling with skip connections within pooling blocks inspired by ResNet \cite{he2016deep}. We find deeper network necessary for learning variegated datasets such as ImageNet. In addition, we include pretrained ImageNet model logits to our layers by concatenation\cite{Liu_2018_ECCV}. We add logits from after pool2, pool3 and pool4 of a VGG model. Finally, to improve the model's ability to learn distant pixels relationships, we add self attention layers proposed by SAGAN in both the encoder and decoder \cite{zhang2018self}. 
\section{Evaluation}
As part of our evaluation, we select three GAN models with different datasets. 
As the baseline, we select improved Wasserstein trained on Dogs vs. Cats dataset to showcase our model's performance for samples with poor quality. We further select StyleGAN and BigGAN as they are cutting edge for GAN models and can demonstrate our approach's strength and weaknesses. StyleGAN is trained on Flickr-Faces-HQ dataset, consisting of 70 000 1024 by 1024 human faces, while BigGAN is trained on 1000 classes of ImageNet. Furthermore, ImageNet contains class labels which we can use to analyze performance per class. Note that for Improved Wasserstein, we train our own model. In contrast, we use pretrained models provided by Nvidia and DeepMind for StyleGAN and BigGAN respectively. 
Given that we are limited by computing resources, we resize training data to 128 by 128 for Flickr, 64 by 64 for ImageNet, and 64 by 64 for Dogs vs. Cats. To train our model, it takes on average 48 hours for the best results. 
Both StyleGAN and BigGAN use a truncation hyper-parameter to control sample quality at the expense of diversity  \cite{karras2018style, brock2018large}.
For truncation, we choose 0.7 for StyleGAN for better sample quality  \cite{karras2018style}. For BigGAN, we use a truncation ranging from 0.1 to 1 sampled uniformly per image. Finally, we use 16 000 samples to compute FID which is sufficient but might lead to worse FID. Due to these factors, our FID score is worse compared to the best scores obtained by both papers. We emphasize this paper is not meant to improve the FID score of the best models but to show our approach can better evaluate images generated from GAN.
\subsection{Evaluation Metrics}
\begin{figure}
  \centering
  \includegraphics[width=14cm]{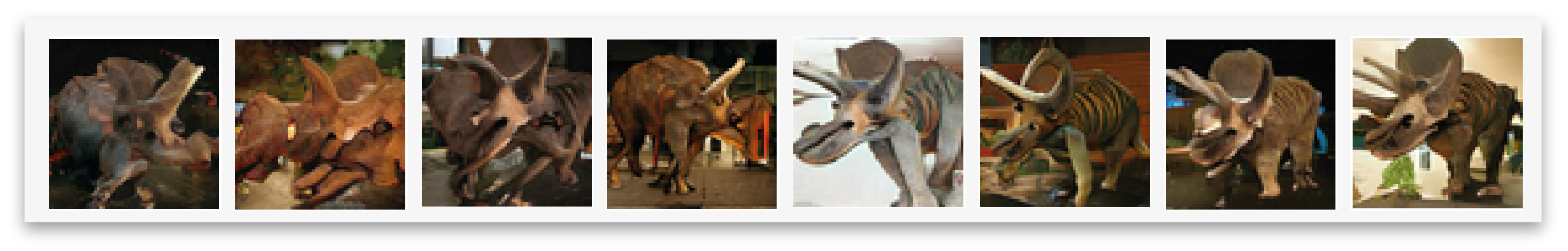}
  \caption{We show the Triceratops class from BigGAN, sorted using PD from worst sample to best sample. We can see an increasing improvement in image quality.}
\end{figure}
\begin{figure}
  \centering
  \includegraphics[width=14cm]{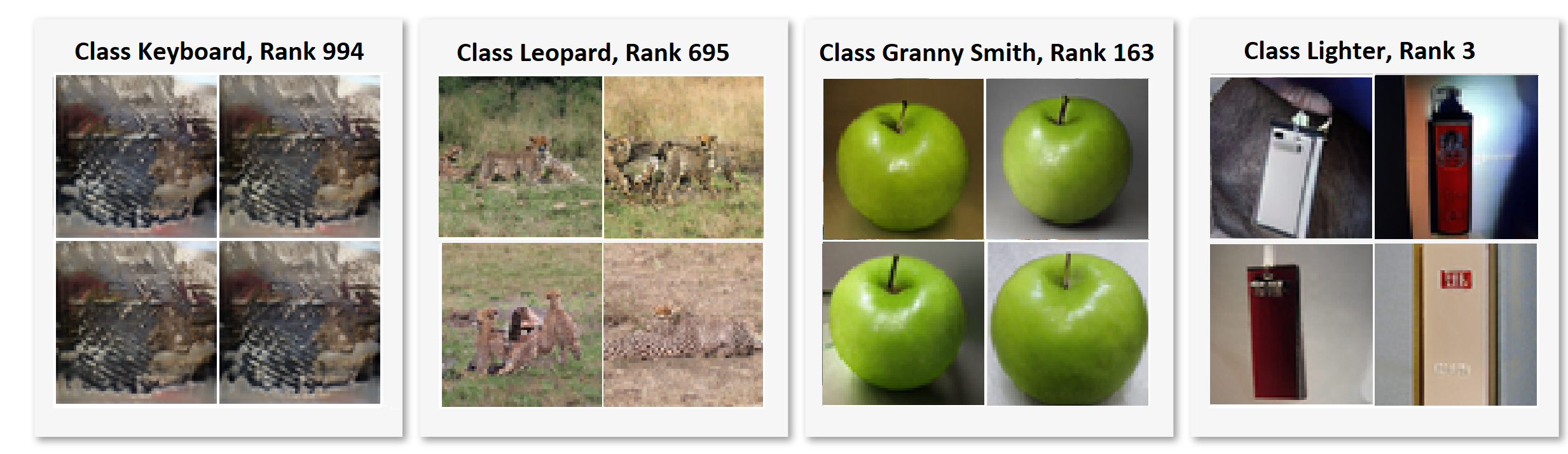}
  \caption{We show ImageNet classes ranked using PD by order of worst to best. For keyboards, we can see both mode collapse and quality issues. Leopards suffers from quality issues and unclear samples. In contrast, granny smiths have high image quality but suffers mode collapse. Finally, the lighters are both high quality and diverse. }
\end{figure}

% \begin{figure}
%   \centering
%   \includegraphics[width=14cm]{32_dog.png}
%   \caption{PD, FID vs Split Number, Dogs and Cat}
% \end{figure}

To evaluate our approach, we rely on both qualitative and quantitative metrics. First, our model can rank individual images, as seen in Figure 3. Furthermore, we can see the difference between classes as ranked by our approach in Figure 4. Second, we also show the PD score for each pixel and its corresponding images as comparison in Figure 1. For mode collapse evaluation, we can use birthday paradox for generated images sampled from a single class. See images in Appendix. 

For quantitative analysis, we show PD correlates well with FID. We sort validation dataset by PD and divide it into splits. Split one contains the worst ranked images and each subsequent split improves in quality. Each individual split's FID is computed against the same real dataset of size 16 000. For ImageNet, we generate the same number of samples per class and sort within each class. The reasoning is that each class is of different quality and mode collapse; directly sorting would cause uneven dataset and lead to strange FID. Finally, we test our model trained on ImageNet for StyleGAN and Improved Wasserstein. For splits of 4, 8, 32, we use validation dataset of size 64 000, 128 000, 512 000. 
\subsection{Training details}
For training, we use AdamOptimizer with a learning rate of 0.0002. For each dataset, 
we must tune hyper-parameters for best results. The value of $\lambda$ is set to 5, $\gamma$ 1, and $L2$ as 0.03. However for Flickr, a $\lambda$ of 2 and $L2$ of 0.3 led to better mode collapse detection. For Dogs vs. Cats, we further add image segmentation and increase the weight of segmented areas in the loss. As shown in Table 1, this approach did not improve results significantly for Dogs vs. Cats and thus, we choose not to use segmentation for Flickr and ImageNet. When training ImageNet, we combine generated images with real images of the same class. Otherwise with 1000 classes, too much variance in the collaged image can prevent the model from learning. This is not necessary for Flickr and Dogs vs. Cats.  
\begin{table}
\begin{center}
 \begin{tabular}{||c c c c c c||} 
 \hline
 Improved Wass, Dogs and Cat & S1 & S2 & S3 & S4 & Random Dataset\\ [0.5ex] 
 \hline\hline
 3 Layer per Pool & 117.7 & 101.9 & 96.4 & \textbf{91.1} & 100.16\\ 
 \hline
 5 Layer per Pool & 123.8 & 105.5 & 95 & \textbf{86.2} & 101.7\\
 \hline
 5 Layer per Pool + Segmentation & 123.1 & 105.2 & 94.5 & \textbf{84.1} & 100.52 \\ [1ex]
 \hline
\end{tabular}
\end{center}
\caption{We show FID results for Dogs vs. Cats samples from improved Wasserstein. A validation dataset of size 64000 was split into 4. }
\label{t1}
\end{table}

\begin{table}
\begin{center}
 \begin{tabular}{||c c c c c c||} 
 \hline
  StyleGAN, Truc=0.7 & S1 & S2 & S3 & S4 & Random Dataset\\ [0.5ex] 
 \hline\hline
 Quality Focused & 25.44 & 16.33 & \textbf{13.61} & 16.56 & 14.37\\ 
 \hline
 Mode Collapse Focused & 30.50 & 17.82 & 12.97 &\textbf{11.04} & 14.06\\ [1ex]
 \hline
\end{tabular}
\end{center}
\caption{We show FID results for StyleGAN and Flickr. A validation dataset of 64000 was split into 4. We tested different values for $\lambda$ and $L2$, one for detecting quality errors and the other for detecting mode collapse. }
\label{t2}
\end{table}

\begin{table}
\begin{center}
 \begin{tabular}{||c c c c c c||} 
 \hline
  BigGAN, 1000 Classes & S1 & S2 & S3 & S4 & Random Dataset\\[0.5ex] 
 \hline\hline
 Trucation Uniform [0.5, 1] & 20.06 & 17.72 & 15.91 & \textbf{14.20} & 17.18\\ 
 \hline
  Trucation Uniform [0.1, 1] & 27.08 & 24.89 & 20.96 &\textbf{16.91} & 21.9\\
 \hline
 Trucation 0.5 & 27.14 & 25.83 & 25.1 & \textbf{24.07} & 25.01\\ 
 \hline
  Trucation 0.4 & 30.07 & 29.02 & 28.4 & \textbf{27.6} & 28.67\\
 \hline
  Trucation 0.3 & 32.77 & 32.41 & 31.8 & \textbf{31.26} & 31.90\\[1ex]
 \hline
\end{tabular}
\end{center}
\caption{We show FID results for BigGAN and ImageNet. Each split consists of 16 images per class for a total of 16 000 per split. Note the FID for real vs real is 4.09 and the model is trained on truncation ranging from [0.1, 1]. }
\label{t3}
\end{table}
\subsection{FID Analysis}
FID as a metric considers both quality and mode collapse. However, when FID score is poor, quality is weighted heavily compared to mode collapse. In contrast, when the FID score is good, mode collapse is weighted heavily over quality. In many cases, we see datasets of images with "higher quality" rated lower than images with clear artifacts because of mode collapse. When we split a dataset into 4 splits, overall FID score for the whole dataset is not the average FID of each individual split. This is due to mode collapse considerations. If we split ImageNet into two groups of 500 classes each, the average FID score will be much higher than the overall FID of the two groups combined. 
\subsection{Improved Wasserstein Analysis}
 \begin{figure*}[t] 
    \centering
  \subfloat[]{%
      \includegraphics[width=5cm,height=4cm]{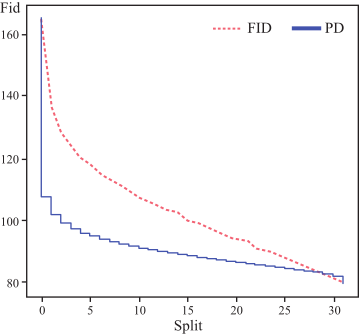}}
      \qquad
  \subfloat[]{%
      \includegraphics[width=6.5cm,height=4cm]{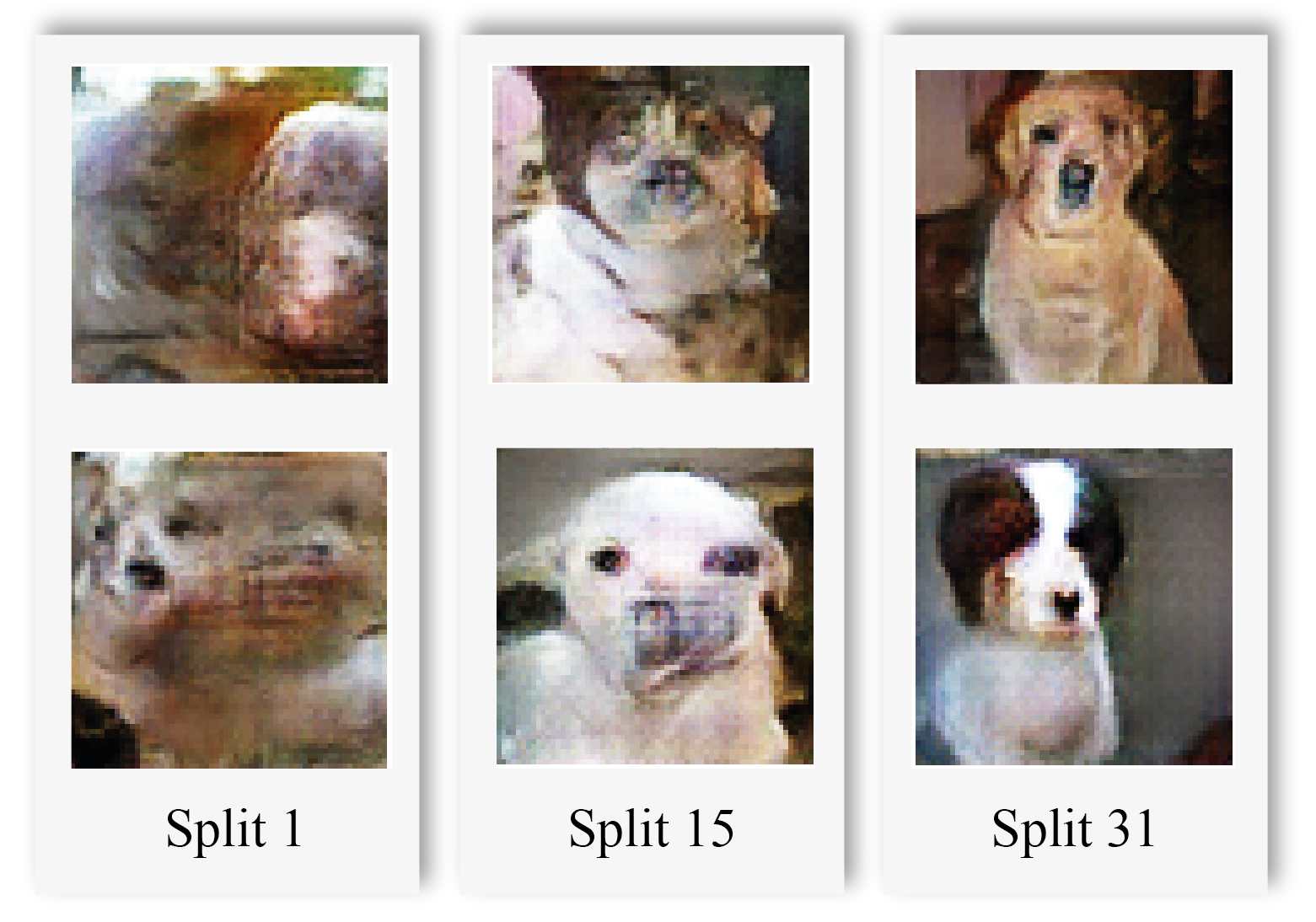}}
%  \includegraphics[width=\textwidth]{ben3.eps}\\
%     \vspace{2cm}
%     \includegraphics[width=\textwidth]{Dogs.png}
      
  \caption{In (a), we evaluate Improved Wasserstein Results by splitting validation into 32 splits ranked by PD. We compare PD with FID and show our score correlates well with FID. For the graph, PD was scaled to match FID. In (b) with the same validation dataset as (a), we selected samples from split 1, split 15, split 31 to show the difference in sample quality.} 
  \label{fig6} 
\end{figure*}

 For Improved Wasserstein, our model performs well for detecting errors as shown in Table 1. Despite improved image quality, there is significant mode collapse for the best rated samples. However, because of low sample quality, our best samples score best for FID. To test the curve of PD and its correlation with FID, we generate 512 000 images and split them into 32 groups. As we see in Figure 5(a), the FID score of the worst group(0) is lower by large margins. This is in part due to mode collapse and bad sample quality from the worst samples. Furthermore, the consistent decrease in FID demonstrates our model is not assigning PD randomly but with purpose. Qualitatively, errors occur on the object while backgrounds tend to be well generated. Samples are shown in Figure 5(b)
\subsection{Style GAN Analysis}
\begin{figure}
  \centering
  \includegraphics[width=13.5cm]{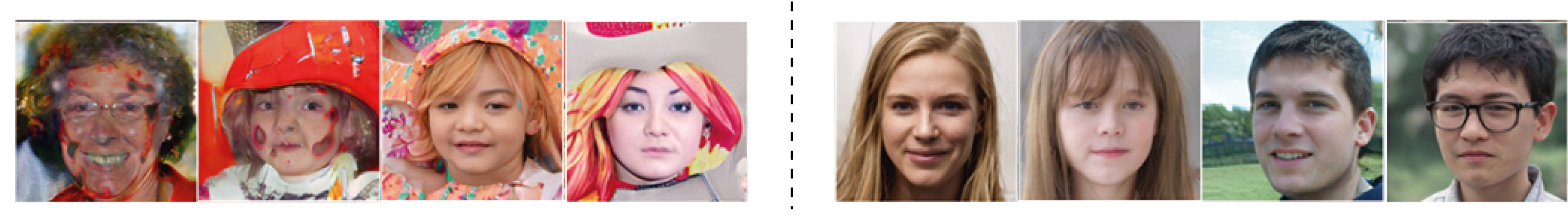}
  \caption{We can see StyleGAN's worst and best samples ranked by our approach}
\end{figure}
The samples from StyleGAN seem to have no errors at first glance. We find instead that StyleGAN still suffers from some quality issues and mode collapse.  By controlling PD with $\lambda$, we can better account for image quality. However, the model prefers bright images which results in a worse FID score for our best ranked split as we see in Table 2. We find it impressive for our model to find badly generated samples from a dataset of such high quality samples, see Figure 1, 6. In contrast, by increasing PD, we can detect mode collapse and show impressive FID performance in Table 2 and appendix. Overall, Style GAN's sample quality is high; there are few samples with severe quality issues. 

\begin{table}[h]
\begin{center}
 \begin{tabular}{||c c c c c c c c c c||} 
 \hline
  Model & S1 & S2 & S3 & S4 & S5 & S6 & S7 & S8 & Random\\[0.5ex] 
 \hline\hline
  Style Truc=0.7 & 22.2 & 16.54 & 14.56 & 13.05 & 11.94 & 11.03 & \textbf{10.27} & 11.49 & 11.61 \\
    \hline
    Impr. Wass. & 108.2 & 105.3 & 102.9 & 101.1 & 99.7 & 97.93 & \textbf{96.94} & 99.12 & 100.2 \\
  [1ex]
 \hline
\end{tabular}
\end{center}
\caption{We use ImageNet trained weights for evaluating other GAN  models. Note that we resize StyleGAN images to 64 by 64 which reduces the FID score by about ~3.}
\label{t4}
\end{table}

\subsection{BigGAN Analysis}
 \begin{figure*}[t] 
    \centering
  \subfloat[]{%
      \includegraphics[width=6.5cm,height=4.5cm]{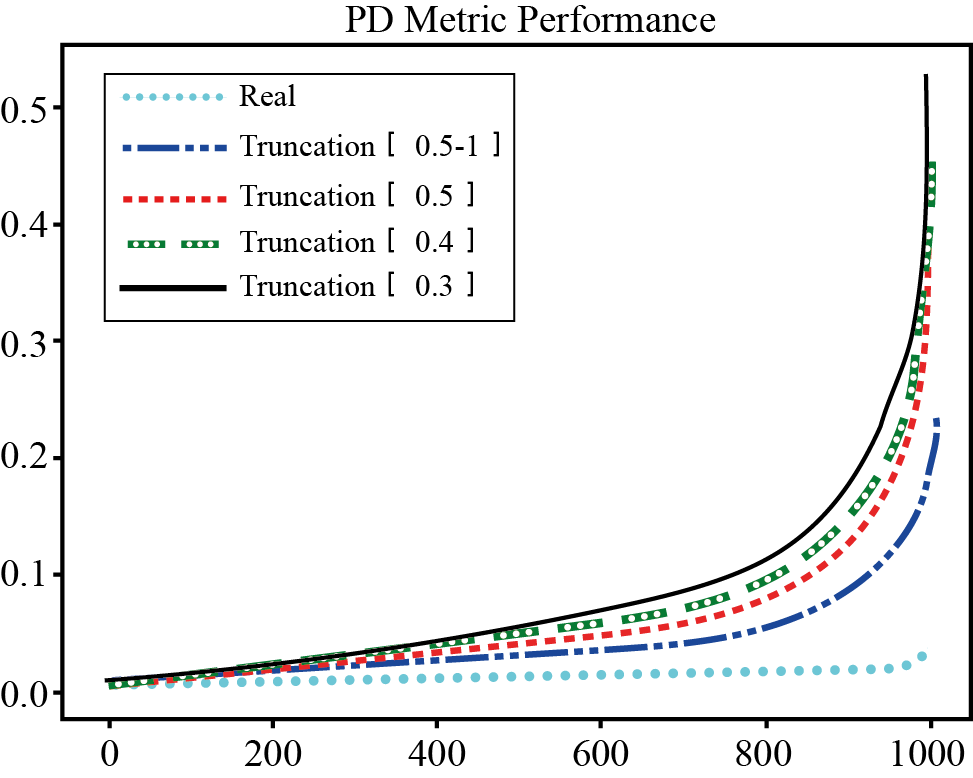}}
      \qquad
  \subfloat[]{%
      \includegraphics[width=6.5cm,height=4.5cm]{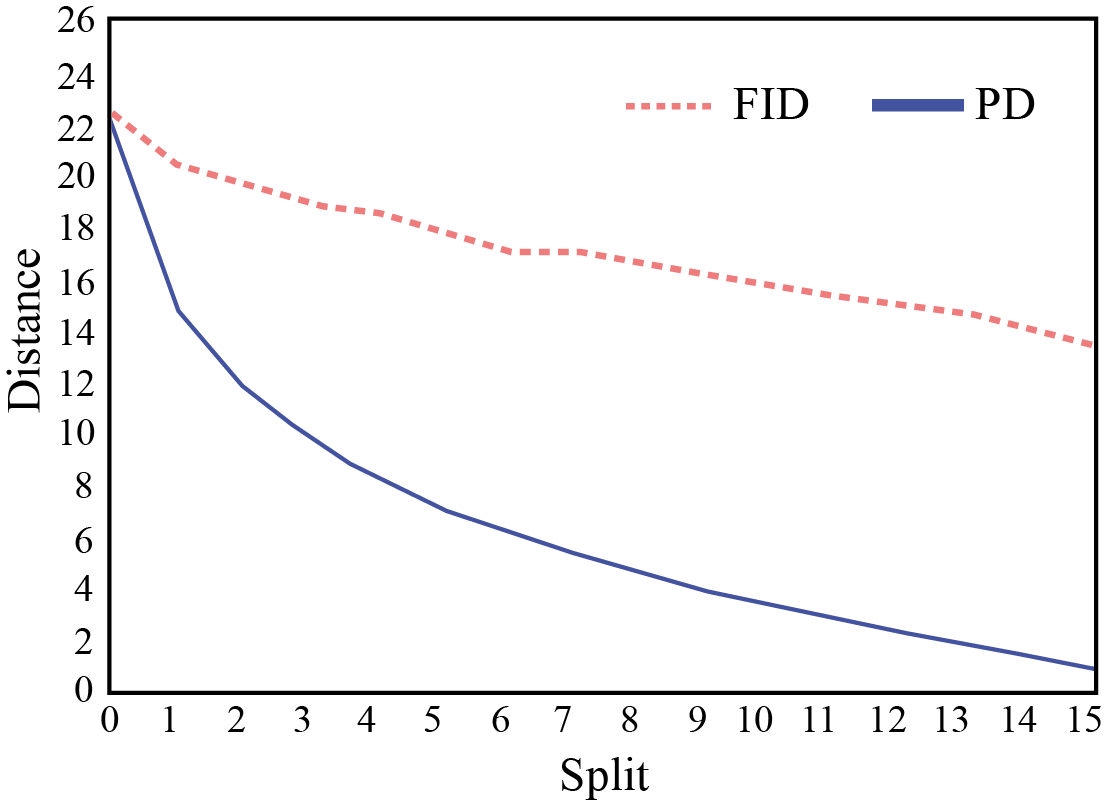}}
  \caption{(a) We graph the sorted PD score for 1000 ImageNet class. Each line consists of different level of truncation. We observe around 200 ImageNet classes suffer from high PD.  
  (b) We further split ImageNet validation with truncation ranging from 0.5 to 1 into 16 splits ranked by PD. We show PD correlates with FID for ImageNet. Note PD is scaled to FID.}
  \label{fig7} 
\end{figure*}
 As seen in Table 3 and Figure 7(b), our results correlates well with FID. Qualitatively, we observe the worse samples suffer from heavy mode collapse and quality issues as seen in appendix. Interestingly, as we add more truncation(more mode collapse), the difference in FID narrows due to reduced differences between images with more mode collapse. Up until now, we have only shown how our metric correlates with FID. However, with PD and labels from ImageNet, we can dig deeper into BigGAN's performance. First with truncation, we show PD correlates with mode collapse as seen in Figure 7(a). Furthermore, we observe that PD score can vary significantly between classes. Samples for different classes can be seen in Figures 3 and appendix. To compare PD score between classes, we offset the generated class's PD by its real class PD. Qualitatively, we observe BigGAN is heavily mode collapsed within each class compared to its real data counterpart which we can observe in the appendix. From Figure 7(a), we see around 200 of BigGAN's classes suffer from high PD. Indeed, a dozen of classes are completely mode collapsed. Observing the best scoring classes, we discover three patterns: comparatively reduced mode collapse within the class, high image quality, and limited diversity from the real class. When we visualize PD for different classes, the location on the image which suffers from high PD vary. For instance, many classes receive high PD in background areas while others such as dogs classes often receive high PD on the face. For certain dogs, BigGAN seems to learn the same face while for snakes, the background is often the same. Finally, we experiment with our pretrained model of ImageNet on StyleGAN and Improved Wasserstein results. Results in Table 4 show that our model can evaluate other GAN models. However the results are not as impressive compared to scores from directly trained models. The model has trouble detecting mode collapse given it was not trained on the data.

\section{Conclusion and Future Work}
In this paper, we show a new approach to evaluate GAN performance for mode collapse and image quality. By combining real and generated images at the pixel level, our approach enables detailed analysis of GAN and can be easily visualized. Furthermore, our approach correlates well with FID and can be used for ranking generated samples. However, our model must directly learn from the generated and real data which requires some training and tuning. The next step would improving its ability for using trained models for other datasets.  
Furthermore, we are looking into using our model directly as a GAN discriminator. Although the stability of the model is unknown, such discriminator could provide more information and better loss to the generator. 
\bibliographystyle{unsrt}  
% \bibliography{nips}

\end{document}